\documentclass[journal,twoside,web]{ieeecolor}
\usepackage{generic}
\usepackage{cite}
\usepackage{booktabs}
\usepackage{amsmath,amssymb,amsfonts}
\usepackage{algorithmic}
\usepackage{graphicx}
\usepackage{algorithm,algorithmic}
\usepackage{hyperref}
\hypersetup{hidelinks=true}
\usepackage{textcomp}
\def\BibTeX{{\rm B\kern-.05em{\sc i\kern-.025em b}\kern-.08em
    T\kern-.1667em\lower.7ex\hbox{E}\kern-.125emX}}
\begin{document}
\title{Comparative Efficacy of Commercial Wearables for Circadian Rhythm Home Monitoring from  Activity, Heart Rate, and Core Body Temperature}
\author{Fan Wu, Patrick Langer, Jinjoo Shim, Elgar Fleisch, and Filipe Barata
\thanks{This work was supported in part by the Scientific Equipment Program of ETH Zurich. \textit {(Corresponding author: Fan Wu.)}}
\thanks{F. Wu, P. Langer, J. Shim, E. Fleisch, and F. Barata are with Centre for Digital Health Interventions, ETH Zurich, Zurich, Switzerland (e-mail: fanwu@ethz.ch; planger@ethz.ch; jshim@ethz.ch; efleisch@ethz.ch; fbarata@ethz.ch). }
\thanks{E. Fleisch is with Centre for Digital Health Interventions, University of St. Gallen, St. Gallen, Switzerland.}
}

\maketitle

\begin{abstract}
Circadian rhythms govern biological patterns that follow a 24-hour cycle. Dysfunctions in circadian rhythms can contribute to various health problems, such as sleep disorders. Current circadian rhythm assessment methods, often invasive or subjective, limit circadian rhythm monitoring to laboratories. Hence, this study aims to investigate scalable consumer-centric wearables for circadian rhythm monitoring outside traditional laboratories. In a two-week longitudinal study conducted in real-world settings, 36 participants wore an Actigraph, a smartwatch, and a core body temperature sensor to collect activity, temperature, and heart rate data. We evaluated circadian rhythms calculated from commercial wearables by comparing them with circadian rhythm reference measures, i.e., Actigraph activities and chronotype questionnaire scores. The circadian rhythm metric acrophases, determined from commercial wearables using activity, heart rate, and temperature data, significantly correlated with the acrophase derived from Actigraph activities (r=0.96, r=0.87, r=0.79; all p$<$0.001) and chronotype questionnaire (r=-0.66, r=-0.73, r=-0.61; all p$<$0.001). The acrophases obtained concurrently from consumer sensors significantly predicted the chronotype ($R^2$=0.64; p$<$0.001). Our study validates commercial sensors for circadian rhythm assessment, highlighting their potential to support maintaining healthy rhythms and provide scalable and timely health monitoring in real-life scenarios.
\end{abstract}


\section{Introduction}
\label{sec:introduction}

\IEEEPARstart{C}{ircadian} rhythms manifest natural 24-hour biological activity patterns~\cite{vitaterna_2001_overview}. Circadian rhythm sleep disorders affect up to 3\% of adults~\cite{kim_2013_circadian}
and sleep duration are associated with 12-30\% elevated risks of mortality~\cite{cappuccio_2010_sleep} and 28-55\% increased incidences of coronary heart disease~\cite{cappuccio_2011_sleep}, type 2 diabetes~\cite{cappuccio_2009_quantity}, and obesity~\cite{cappuccio_2008_metaanalysis}. Continuous monitoring of circadian rhythm allows personalized adjustments to daily routines. For example, optimizing lighting schedules can facilitate circadian adaptation to shifted sleep and wake schedules~\cite{rahman_2022_dynamic}.

Melatonin, core body temperature (CBT), and rest-activity rhythm profiles are commonly used to measure circadian timing~\cite{reid_2019_assessment}. However, these measurements are primarily conducted in labs using invasive techniques like blood serum sampling for melatonin and rectal thermometry for CBT, limiting their use only in controlled laboratory settings. In contrast, Actigraph, a wrist-worn accelerometer, is commonly used to assess activity profiles as well as sleep and awake patterns. Offering an objective depiction of an individual's rhythm patterns, Actigraph has emerged as the reference method for circadian rhythm measurement~\cite{ancoliisrael_2003_the, ali_2023_circadian}. However, it lacks instant data access due to limited connectivity options and requires a closed-source software license for data access, hindering continuous monitoring. Chronotype questionnaires like the Morning-Eveningness Questionnaire (MEQ) have proven to be valid for predicting intrinsic circadian period and sleep patterns but demand active involvement for sample collection~\cite{horne_1976_a, taillard_2004_validation}. The limitations of each method pose challenges to scaling up for widespread population use, underscoring the need for improved accessibility and usability solutions.

Alternative to traditional laboratory tests, the rapid advancement of wearables has driven the development of digital biomarkers, offering a practical and convenient solution for real-life tracking of circadian rhythms. Wearables like smartwatches, measuring parameters like acceleration, heart rate (HR), and heart rate variability (HRV), could enable scalable rhythm monitoring at the population level due to their widespread use. These obtained biomarkers facilitate real-time rhythm analysis with minimal user involvement, supporting personalized, preventative disease management through self-monitoring in everyday contexts. Focusing on commercially available devices ensures accessibility to a wide user base, potentially aiding in population-wide circadian rhythm monitoring and the development of digital health interventions aimed at enhancing treatment outcomes.

Despite the acknowledged potential of wearable technology, there is limited empirical evidence validating its efficacy in studying circadian rhythms, especially with commercially available devices. Using a clinical-grade sensor, Suarez et~al. predicted circadian phase using one week of ambulatory activity and HR data~\cite{suarez_2021_circadian}. Regrettably, due to its high cost and the closed-license software, it cannot facilitate longitudinal measurements for a broader population. In contrast, several studies focused on more widely available commercial wearables, such as smartwatches. Nevertheless, existing research aiming to assess activity patterns using smartwatches often faces limitations such as a lack of continuous measurements~\cite{lee_2017_comparison}, absence of direct assessments regarding device accuracy~\cite{huang_2021_predicting}, or constraints due to small sample sizes~\cite{smagula_2021_initial}. Fewer studies collect other data like CBT and HR to assess circadian rhythm using wearables. Due to challenges in measuring CBT outside clinical settings, researchers have turned to alternatives like iButton for skin temperature, showing a negative correlation with activity ($r=-0.56$)~\cite{vanmarkenlichtenbelt_2006_evaluation}. In HR measurement with the Apple Watch, circadian phase estimation differs from melatonin lab results by 4.4 hours\cite{huang_2021_distinct}. These findings suggest a need for further evaluation of commercial wearables for circadian rhythm assessment, particularly for longitudinal data, essential for home monitoring. 

Challenges in evaluating commercial wearables persist, including achieving stable data collection and transmission from home settings, validating the accuracy and feasibility of wearable sensors, and addressing user acceptance, including concerns regarding device charging. Additionally, the lack of standardized and open-sourced methods further complicates comparisons between different techniques and devices.

This paper aims to validate the efficacy of commercial devices in assessing circadian rhythm longitudinally in real-world settings. Therefore, we conducted an observational longitudinal study to concurrently monitor activity, CBT, and HR using commercially available sensors for circadian rhythm assessment at home. Due to the difﬁculties in measuring melatonin from home, we chose the MEQ questionnaire and ActiGraph as our references for comparing efficacy. We explored practical implications for real-world monitoring, including data collection stability and wearing comfort. Accounting for real-life factors like charging time and missing data, we evaluated the accuracy of consumer sensors by comparing their circadian rhythm metrics with those obtained from Actigraph and MEQ scores. Monitoring with these commercial devices can help individuals understand their body clock and maintain a healthy rhythm. 

To promote standardized procedures in circadian rhythm research, we developed an application to collect smartwatch data using our platform CLAID. We release our software and packages as open-source. The data analysis and collection software packages can be accessed via [link to be provided] and https://claid.ethz.ch. Additionally, the datasets created in the study are available in [link to be provided].

\section{Materials and Methods}

\subsection{Subjects and Protocol}

We recruited 38 healthy subjects between May 23rd, 2023, and October 5th, 2023. Subjects aged 18 to 65, with no prior history of allergic contact dermatitis, unstable concurrent diseases, specific ICD-10 conditions affecting circadian rhythms (e.g., sleep disorders, psychiatric and rheumatic diseases), debilitating mental illnesses (e.g., dementia, Alzheimer's), or cardiovascular events, were expected. Additionally, pregnant women were not considered. Before commencing the study, all subjects signed written informed consent forms outlining the study's details and privacy regarding personal information. During the 14-day longitudinal observational study, each subject wore three sensors continuously for 24 hours each day: an ActiGraph, a smartwatch, and a CBT sensor. Besides, participants completed the MEQ questionnaire thrice on Days 0, 7, and 14. This study was approved by the Cantonal Ethics Commission of Zurich (BASEC No. 2023-00397).

Two participants discontinued the study due to skin problems. Consequently, 36 successfully participated in the study, including 17 (47\%) females. The participant’s age was $33\pm10.02$ years (range 23–58), and BMI was $22.6\pm2.2$ kg/m$^{2}$ (range 19.4–28.4). The sample was ethnically diverse: 77.8\% White, 13.89\% Asian, and 8.3\% Hispanic or Latino. Twenty participants (55\%) were students enrolled in the university, and the rest worked full-time. Their work or study hours varied: 7 participants (19.44\%) over 50 hours per week, 6 (16.67\%) between 45-50, 15 (41.67\%) between 40-45, 4 (11.11\%) between 35-40, and rest between 30-35. 

\subsection{Device and Data}

For continuous CBT monitoring, we chose the CALERAresearch sensor (GreenTEG, Switzerland). It accurately predicts CBT by merging skin temperature data and heat flux signals using a clinically validated algorithm~\cite{ajevi_2022_a,rerabek_2022_circadian}, unlike other solutions limited to skin temperature measurement such as iButton or TempTraq~\cite{flora_2021_highfrequency}. The GreenTEG-developed Android application, ``CORE Body Temperature Monitor'' (Firmware version 0.8.0), collected temperature data every minute on the participant's smartphone and then transferred data to its server. From the server, we acquired skin temperature (SkinT) and CBT data at the minute level. The sensor assigned signal quality scores ranging from 1 (poor) to 4 (best). We retained data with a quality score of 2 or higher.

We selected the Galaxy Watch5 (Samsung Electronics, South Korea) for our study, considering its cost, battery life, data storage, SDK availability, and development effort. Using the Samsung privileged health SDK, we developed a study application to collect data through our CLAID platform. CLAID is an open-source, cross-platform framework for sensor integration and data collection over various devices and operating systems~\cite{langer_2023_claid}. Data was securely transmitted to an ETH Zurich server during smartwatch charging. In connection loss, data was temporarily stored on the smartwatch and synchronized upon reconnection. Daily server checks ensured data integrity, and questionnaires were emailed via the server. Data was uploaded to Leonhard Med Secure Scientific Platform for storage and processing upon study completion.

The smartwatch collected 3-axis raw acceleration and HR-related data at a sampling rate of 25 Hz. HR-related data included HR and R-R intervals (RRI), i.e., R-wave peak to R-wave peak in electrocardiograms, representing the interval between consecutive heartbeats. We segmented RRI data into 10-minute intervals and calculated four time-domain HRV metrics: Mean RR (mean of RRI), SDNN (standard deviation of RRI), RMSSD (root mean square of successive RRI differences), and pNN50 (percentage of successive RRI differing by more than 50ms).

The ActiGraph wGT3X-BT (ActiGraph, USA), sampling acceleration data at 80 Hz, stored data locally on the device. Upon participants' device return, we obtained Activity Counts (ACs) using ActiGraph's proprietary software that are calculated as the cumulative sum of post-filtered accelerometer values within one-minute epochs~\cite{neishabouri_2022_quantification}. We adopted this algorithm and derived the smartwatch’s activity counts (WatchAC) from its raw acceleration data. To evaluate the smartwatch's accuracy, we compared WatchAC with ACs obtained from ActiGraph software (ActiAC), serving as our reference. We aligned temporal data by identifying sleep time and excluding non-wear times for both devices, including smartwatch charging and outlier data while retaining overlapping usage periods. Analyses were conducted using Python 3.9.16 and R 4.1.3.

\subsection{Circadian Rhythm Metrics}

Based on established techniques, we interpreted circadian rhythm using cosinor parameters and non-parametric measures~\cite{bate_2023_the}. The cosinor method, commonly used in previous research, aligns data with a cosine curve, as shown in \eqref{eq}, providing insights into fundamental circadian functions like sleep timing~\cite{dawes_1972_circadian, spengler_2000_endogenous}. In our study, we assumed a 24-hour period for the curve.

\begin{equation}
y(t)=M+Acos(\frac{2\pi t}{\tau}+\varphi)
\label{eq}\end{equation}

$y$ represents the sensor data, i.e., activity, temperature, and HR-related data. Circadian rhythm can be interpreted by three cosinor parameters: the mesor ($M$), representing the mean of the modeled rhythm; the acrophase ($\varphi$), indicating the time at which the fitted curve reaches its peak; and the amplitude ($A$), signifying half the extent of predictable variation within a single cycle. Aggregating sensor data into 10-minute intervals, we applied a one-component cosinor model to fit 14 days of data into the 24-hour rhythm, yielding cosinor parameters for each sensor data type~\cite{mokon_2020_cosinorpy}. 

However, the cosinor model cannot measure rhythm fragmentation, which refers to the disruption of a regular rhythm due to daytime naps or nocturnal activity episodes~\cite{fossion_2018_quantification}. To comprehensively assess circadian rhythm, we computed non-parametric measures for each sensor data type, including intradaily variability (IV), quantifying rhythm fragmentation; interdaily stability (IS), estimating the consistency of activity patterns across multiple days; the average activity during the least active 5-hour period (L5); the average activity during the most active 10-hour period (M10); and relative amplitude (RA), with a higher value indicating a balanced rest-active status.

To summarize, we considered eight types of sensor data from commercial wearables: WatchAC, CBT, SkinT, HR, and four HRV measures, i.e., Mean RR, SDNN, RMSSD, and pNN50. For each type of sensor data, we calculated eight circadian rhythm metrics, including cosinor parameters such as mesor, acrophase, and amplitude, along with non-parametric measures like IS, IV, M10, L5, and RA. We applied Min-Max normalization to the sensor data to compare amplitude and mesor across different sensor data. Analyses were processed using Python 3.9.18. 

\subsection{Chronotype Prediction and Group}

We used the chronotype questionnaire MEQ as another reference to assess participants' circadian rhythm in home settings. We calculated the MEQ scores of each participant by averaging three evaluations conducted over two weeks. Since acrophase represents the temporal order of the human time structure~\cite{ticher_1994_human}, we employed linear regression models to evaluate how well the acrophase of sensor data, including ActiAC, WatchAC, CBT, SkinT, HR, Mean RR, and RMSSD, could predict circadian rhythm. MEQ was the dependent variable in the models, while the acrophases acted as independent variables (features or predictors). The omnibus test confirmed the normal distribution of errors in all models. 

First, we predicted MEQ scores using single features. Then, with multiple regression models, we aimed to enhance circadian rhythm assessment by leveraging the most relevant combination of features from multiple sensors. With a focus on assessing consumer sensors, we excluded sensor data from Actigraph (ActiAC) in the feature combinations. We selected the top $k$ features ($k$ ranging from 2 to 6), explaining MEQ variations most effectively. We assessed multicollinearity in the multiple regression models by calculating the Variance Inflation Factor (VIF) to account for potential high correlation among features. For the sensitivity analysis, age and sex (encoded as 0 for females and 1 for males) served as control variables in the regression models. 

Furthermore, we categorized the entire population into three distinct groups based on individuals' MEQ scores: Evening (scores below 42), Morning (scores exceeding 58), and Intermediate (remaining individuals)~\cite{zacharia_2014_development}. We explored whether acrophase derived from sensor data could elucidate the differences among these groups. The sensor data included ActiAC, WatchAC, CBT, SkinT, HR, Mean RR, and RMSSD.

\subsection{Statistical Analysis and Evaluation}
We assessed normality and described normal distributions as mean$\pm$SD and non-normal distributions as medians with interquartile ranges (IQRs). When applicable, we reported the range (minimum-maximum) for continuous variables or proportions for categorical variables.

%

\subsubsection{Assess Smartwatch Acceleration Accuracy}
We compared WatchAC with ActiAC for each participant using Pearson's correlation and its respective p-value $r(p)$. Additionally, we performed a Bland-Altman analysis to assess the agreement between WatchAC and ActiAC by comparing their differences to the average of the two ACs. Furthermore, we calculated a repeated measures correlation ($r_m$) with its degree of freedom, p-value, and 95\% confidence intervals ($95\%CI$), addressing within-participant correlations by considering paired measurements across multiple participants.

\subsubsection{Assess Commercial Devices for Rhythm Metrics}
Using ActiAC as the reference, we assessed how well commercial wearables interpreted circadian rhythm. We calculated Pearson's correlation and its respective p-value $r(p)$ between the circadian rhythm metrics derived by commercial wearables and those calculated from ActiAC. Since the data did not conform to a normal distribution ($p>0.05$) according to the Shapiro-Wilk test, we chose the non-parametric Wilcoxon signed-rank test to compare the differences in rhythm metrics calculated from commercial wearables and ActiAC. A significant p-value in Wilcoxon signed-rank test $p_{wst}<0.05$ indicates a significant difference between compared metrics. 

\subsubsection{Assess Acrophase-Predicted Chronotype Score}
We calculated the association between MEQ scores and acrophases from Actigraph and commercial sensors using Pearson's correlation $r(p)$. We then conducted a pairwise correlation analysis to examine the relationship between the predicted MEQ score and its predictors. 

\subsubsection{Assess Acrophase Variation Across Chronotype Groups}
Since the acrophase derived from sensor data did not follow a normal distribution ($p>0.05$), we used non-parametric tests, i.e., the Kruskal-Wallis test, to identify differences among at least two groups. Additionally, we employed the Wilcoxon rank-sum test for pairwise comparisons among three chronotype groups to identify specific group differences.

\subsection{Data Transmission Rate and Device Comfort}
We computed data miss rates of three worn sensors for each participant, indicating the percentage of unrecorded data during collection periods. Participants were queried about missed data instances, and reasons were reported. To assess device comfort, participants rated the comfort of two consumer devices, i.e., the smartwatch and the CBT sensor, on a scale from 1 (not comfortable at all) to 5 (very comfortable). They were asked about their willingness to wear these commercial devices and to specify the monitoring frequency.

\section{Results}

\subsection{Assess Smartwatch Acceleration Accuracy}

\begin{figure*}[!t]
    \centering
    \includegraphics[width=1\textwidth]{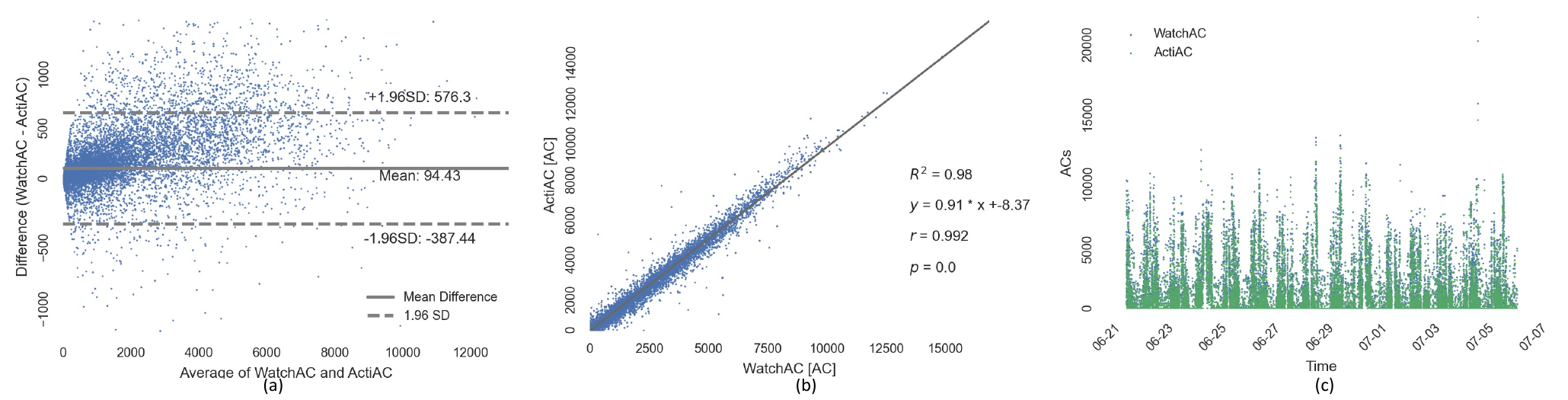}
    \caption{Comparison between WatchAC and ActiAC of one individual. (a) Bland-Altman plot showing the agreement between two sets of ACs. (b)~Regression line with $R^2$ and Pearson’s coefficient with p-values. (c)~Scatter plot of WatchAC (blue) and ActiAC (green) for two weeks.}
    \label{fig1}
\end{figure*}

WatchAC and ActiAC showed a strong correlation over the two weeks ($r$ range 0.894-0.996; all $p<0.001$; N=36). An average correlation of $r=0.984\pm0.019$ implied that the correlations are relatively consistent among participants. The repeated measures correlation of ActiAC and WatchAC over all participants during two weeks was significantly strong ($r_m(619613)=0.982$, $p=0.0$, $95\%CI=[0.98,0.98]$). 
For temporal alignment, non-wearing periods were excluded, removing an average of 1.49\% of data (median: 0.88\%) for both devices not worn and 0.49\% (median: 0.31\%) for single-device non-wearing periods across all participants. Using one participant as an example, as shown in Fig.~\ref{fig1}, the mean difference between WatchAC and ActiAC of 94.43 ACs indicated the Smartwatch overestimated ACs compared to Actigraph. No apparent increase in the differences between WatchAC and ActiAC was noted at higher activity levels. 

\subsection{Assess Commercial Devices for Rhythm Metrics}

Table.~\ref{tab1} displays Pearson's correlation coefficients with their p-values $r(p)$ and Wilcoxon signed-rank test's p-values $p_{wst}$, comparing eight circadian metrics derived from ActiAC with those computed from commercial sensors. For activity data, no significant ($p_{wst}>0.05$) difference was observed between WatchAC and ActiAC across a variety of circadian metrics, i.e., acrophase (15.78 vs. 15.74), IV (0.45 vs. 0.45), M10 (0.09 vs. 0.09), L5 (0.01 vs. 0.01) and RA (0.89 vs. 0.89). Correlations between WatchAC and ActiAC were over 0.89 ($p<0.001$) for all circadian metrics. For temperature data, though CBT demonstrated a significant ($p_{wst}<0.001$) difference in all circadian metrics with ActiAC, its acrophase (16.6 vs. 15.74, $p<0.001$) and mesor (0.33 vs. 0.09, $p<0.01$) were significantly correlated with ActiAC. SkinT significantly differed from ActiAC in most circadian metrics except for amplitude (0.08 vs. 0.07). Its acrophase significantly correlated with ActiAC ($r=-0.40$, $p<0.05$). For heart data, HR was significantly different from ActiAC in all metrics ($p_{wst}<0.01$). Its acrophase and IS significantly correlated with ActiAC ($r=0.87$, $r=0.54$; both $p<0.001$). As for the four HRV measures, most circadian metrics of those measures were significantly different from those of ActiAC ($p_{wst}<0.05$), except for SDNN’s amplitude and IS with pNN50’s amplitude and IV. Acrophase of MeanRR was most significantly correlated with that of ActiAC ($r=0.79$, $p<0.001$).

\begin{table*}[!t]
    \centering
    \caption{Circadian rhythm metrics calculated from consumer devices compared to those derived from Actigraph}
    \begin{tabular}{l|c|cccccccc}
        \hline
        \hline
        \toprule
        \textbf{Metrics} & \textbf{Reference} & \multicolumn{8}{c}{\textbf{Commercial Sensor Data}} \\
         & ActiAC & WatchAC & CBT & SkinT & HR & MeanRR & SDNN & RMSSD & pNN50 \\
        \hline
        \midrule
        \textbf{Amplitude} & 0.07(0.04) & 0.06(0.04) & 0.13(0.07) & 0.08 (0.05) & 0.1(0.05) & 0.16(0.05) & 0.08(0.06) & 0.11(0.04) & 0.06(0.05) \\
        $r(p)$ & & \textbf{0.89***} & 0.14 & 0.05 & \textbf{0.39*} & 0.16 & -0.03 & -0.02 & -0.10 \\
        $p_{wst}$  & & {0.04} & {0} & 0.38 & {0} & {0} & 0.26 & $<$0.01 & 0.67 \\
    \textbf{Acrophase[h]} & 15.74(1.61)	& 15.78(1.96) & 16.6(2.16)	& 3.0(2.65)	& 15.81(1.73)	& 4.06(1.51)	& 14.51(2.77)	& 15.21(1.79)	& 14.99(1.36)\\
        $r(p)$ & & \textbf{0.96***}	& \textbf{0.79***}	& \textbf{-0.40*}	& \textbf{0.87***}	& \textbf{0.79***}	& 0.08	& \textbf{0.35*}	 & \textbf{0.36*} \\
        $p_{wst}$ & & 0.29	& $<$0.001 & 0	& $<$0.01 &	0	& 0	& $<$0.01 &	$<$0.01 \\
    \textbf{Mesor} & 0.09(0.06)	& 0.08(0.06)	& 0.33(0.1)	& 0.63(0.07)	& 0.27(0.11)	& 0.51(0.07)	& 0.35(0.08)	& 0.29(0.06)	& 0.58(0.08) \\
        $r(p)$ & & \textbf{0.93***}	& \textbf{0.46**}	& 0.07	& 0.27	& -0.18	& 0.12	& 0.09	& 0.09 \\
        $p_{wst}$ & & 0	& 0 & 0 &	0	& 0 &	0 &	0	& 0 \\
    \textbf{IS} & 0.24(0.08) &	0.24(0.06) &	0.42(0.19) &	0.25(0.21) &	0.39(0.18) &	0.41(0.13)	& 0.23(0.13)	& 0.3(0.17)	& 0.57(0.41) \\
        $r(p)$ & & \textbf{0.98***} &	0.19 &	\textbf{0.38*}	& \textbf{0.54***}	& -0.19 &	-0.27	& -0.20 &	-0.13 \\
        $p_{wst}$ & & $<$0.001	& 0	& 0.03 &	0	& 0	& 0.59 &	0.03	& 0 \\
        \textbf{IV} & 0.45(0.09) &	0.45(0.09) &	0.0(0.0)	& 0.01(0.01) &	0.01(0.0) &	0.22(0.08)	& 0.74(0.22)	& 0.65(0.27)	& 0.4(0.31)\\
        $r(p)$ & & \textbf{0.99***}	& 0.004 &	0.13	& \textbf{0.43**}	& 0.02	& 0.18	& 0.05	& -0.21 \\
        $p_{wst}$ & & 0.11	& 0 &	0	& 0	& 0	& 0	& 0	& 0.44 \\
        \textbf{M10} & 0.09(0.03)	& 0.09(0.03)	& 0.42(0.11)	& 0.73(0.1) &	0.35(0.1)	& 0.51(0.07)	& 0.35(0.08) &	0.29(0.06)	& 0.58(0.1) \\
        $r(p)$ & & \textbf{0.98***} &	0.31	& -0.002	& 0.21	& -0.07 &	-0.08 &	-0.02 &	-0.16 \\
        $p_{wst}$ & & 0.18	& 0	& 0	& 0	& 0	& 0 & 0 & 0 \\
        \textbf{L5} & 0.01(0.0)	& 0.01(0.0) &	0.17(0.08)	& 0.56(0.16)	& 0.18(0.08)	& 0.51(0.07) &	0.35(0.08)	& 0.29(0.06)	& 0.58(0.1) \\
        $r(p)$ & & \textbf{0.99***} &	0.27 &	0.20	& 0.15	& -0.29 &	0.04	& 0.19	& -0.07 \\
        $p_{wst}$ & & 0.3	& 0	& 0	& 0	& 0	& 0 & 0 & 0 \\
        \textbf{RA} & 0.89(0.07)	& 0.89(0.07)	& 0.41(0.13)	& 0.12(0.1)	& 0.31(0.12)	& 0.0(0.0)	& 0.0(0.0)	& 0.0(0.0)	& 0.0(0.0) \\
        $r(p)$ & & \textbf{0.99***}	& 0.03	& \textbf{0.36*}	& 0.06	& 0.002 &	-0.07	& -0.08	& -0.11 \\
        $p_{wst}$ & & 0.29	& 0	& 0	& 0	& 0	& 0 & 0 & 0 \\
        \hline
        \hline
        \bottomrule
    \end{tabular}
    \label{tab1}
    \vspace{0.2cm}
    \parbox{\linewidth}{\raggedright Each metric is reported as medians (IQRs). Significant levels of correlation are denoted in bold: $p<0.001$(***), $p<0.01$(**), and $p<0.05$(*). }
\end{table*}

\subsection{Assess Acrophase-Predicted Chronotype Score}

Participants' chronotype MEQ scores, $52.86\pm10.61$ with a range of 30.67-74.00, exhibited significant inverse correlations with the acrophase of ActiAC, HR, Mean RR, WatchAC, and CBT ($r=-0.71$, $r=-0.73$, $r=-0.71$, $r=-0.66$, $r=-0.61$, all $p<0.001$). The acrophase of RMSSD exhibited a moderate inverse correlation with MEQ ($r=-0.36$, $p<0.05$), while SkinT was the only variable showing a positive correlation with MEQ ($r=0.38$, $p<0.05$). No significant correlations existed between MEQ and the acrophases from SDNN and pNN50 ($r=-0.11$, $r=0.01$).

\begin{figure}[!t]
\centerline{\includegraphics[width=0.95\columnwidth]{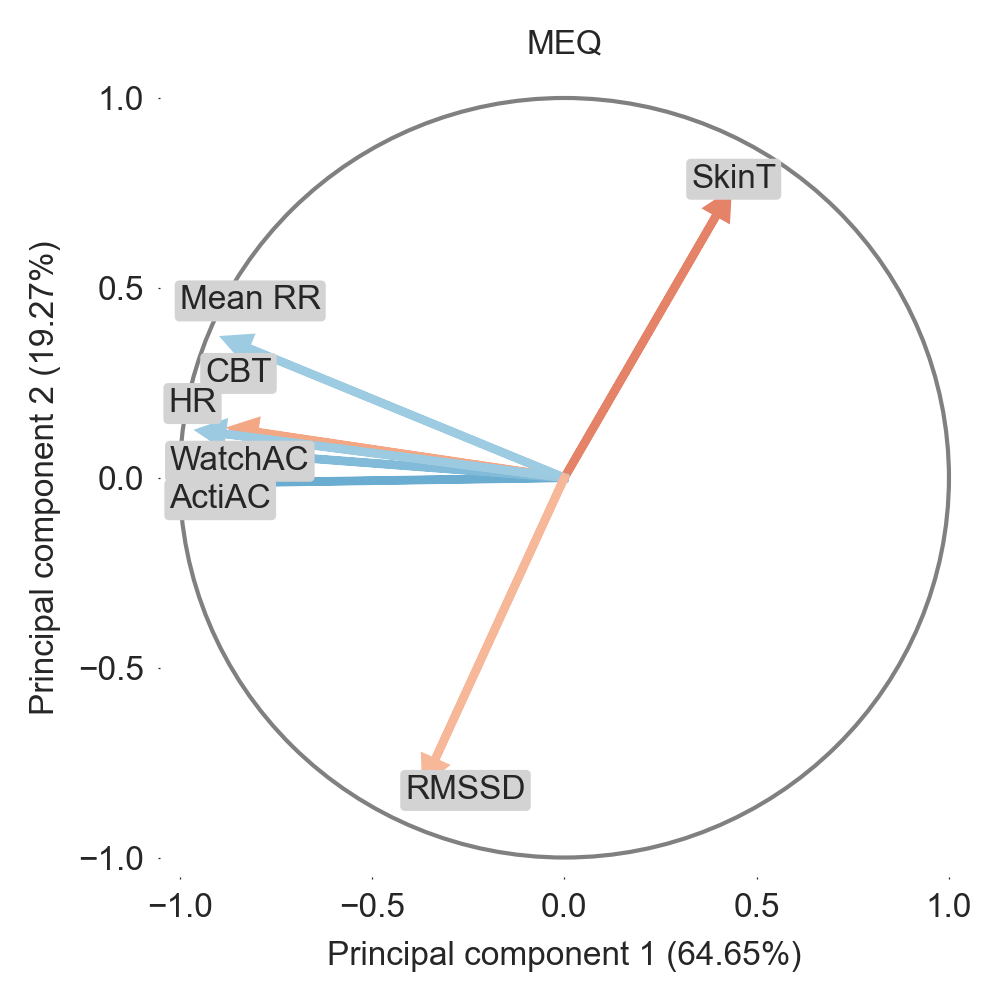}}
\caption{The biplot shows the relationship between MEQ scores and sensor data acrophase. Vectors represent the strength and direction of their correlation on the first and second principal components.}
\label{fig2}
\end{figure}

We predicted MEQ using acrophase from the sensor data with linear regression models. Table.~\ref{tab2} shows regression coefficients $\beta$ with [$95\% CI$], $F$-statistics and $R^2$ of each model. Single linear regression models (models 1-7) indicated statistical significance ($p<0.05$). Except for the acrophase of SkinT and RMSSD (models 4 and 7), all other predictors explained over 35\% of the variance in MEQ scores. The acrophase of HR was the most significant negative predictor of MEQ scores ($\beta=-6.36$, $R^2=0.53$, $F(1,34)=38.22$, $p<0.001$), closely followed by the ActiAC ($R^2=0.51$, $F(1,34)=34.76$, $p<0.001$) and Mean RR ($R^2=0.51$, $F(1,34)=34.63$, $p<0.001$). SkinT was the only positive predictor ($\beta=0.56$, $R^2=0.14$, $F(1,34)=5.58$, $p<0.05$). As illustrated in Fig.~\ref{fig2}, we drew a principal component analysis biplot of MEQ scores with acrophase of ActiAC, WatchAC, CBT, SkinT, HR, MeanRR, and RMSSD as vectors. Most of the variation was explained by the first two principal components (PCs) (83.92\%), with the first PC having the highest contribution of 64.65\%.

\begin{table}
\caption{Linear regression predicting MEQ scores using various acrophase as predictors}
\label{table}
\setlength{\tabcolsep}{3pt}
    \begin{tabular}{l|cccccc}
        \hline
        \hline
        \toprule
        \textbf{Model} & \textbf{Predictors} & \textbf{$\beta$ [95\%CI]} & \textbf{VIF} & \textbf{$F$} & \textbf{$R^{2}$} & \textbf{Adj $R^{2}$} \\
        \hline
        \midrule
        \textbf{1} & Acti & \textbf{-5.27[-7.09 -3.45]***} & & \textbf{34.76***} & 0.51  \\
        \textbf{2} & WatchAC	& \textbf{-4.43[-6.17 -2.69]***} & &	\textbf{26.80***}	&0.44 \\
        \textbf{3} & CBT &	\textbf{-3.64[-5.28 -2.01]***}	& &	\textbf{20.48***} &	0.38 \\
        \textbf{4} & SkinT	& \textbf{0.56[0.08 1.05]*}	& &	\textbf{5.58*}	& 0.14 \\
        \textbf{5} & HR	& \textbf{-6.36[-8.45 -4.27]***}	& &	\textbf{38.22***} & 0.53  \\
        \textbf{6} & Mean RR & \textbf{-6.31[-8.48 -4.13]***} & &	\textbf{34.63***} &	0.51  \\
        \textbf{7} & RMSSD &	\textbf{-1.74[-3.32 -0.16]*}	& &	\textbf{4.98*} &	0.13 \\
        \textbf{8} & HR	& -4.29[-9.92 1.33]	& 7.15	& \textbf{19.24***} &	0.54	& 0.51 \\
                 & Mean RR &	-2.26[-7.98 3.45] &	7.15  \\
        \textbf{9} & WatchAC	& -0.93[-4.07 2.21] &	3.75	& \textbf{12.70***} &	0.54	& 0.50  \\
                 & HR	& -3.39[-9.84 3.05]	& 9.17  \\
                 & Mean RR	& -2.10[-7.90 3.70]	& 7.21  \\
        \textbf{10} & WatchAC	& -0.36[-3.98 3.27]	& 4.89 &	\textbf{9.48***} &	0.55& 0.49  \\
                 & CBT	& -0.81[-3.29 1.66]	& 2.87  \\
                 & HR	& -3.36[-9.87 3.15]	& 9.18  \\
                 & Mean RR &	-1.89[-7.78 4.01] &	7.30  \\
        \textbf{11} & WatchAC	& 0.11[-3.41 3.64] 	& 4.99	& \textbf{8.89***} &	0.60	& 0.53 \\
                 & CBT	& -0.49[-2.90 1.91]	& 2.93  \\
                 & SkinT & 0.38[-0.04 0.79]	& 1.35  \\
                 & HR	& -1.51[-8.10 5.08] &	10.13 \\
                 & Mean RR	& -4.22[-10.44 2.01]	& 8.77  \\
        \textbf{12} & WatchAC	& 0.54[-2.87 3.94]	& 5.08	&\textbf{8.75***}	& 0.64	& 0.57 \\
                 & CBT	& -0.80[-3.13 1.53]	& 2.99  \\
                 & SkinT & 0.26[-0.15 0.68]  &	1.48  \\
                 & HR &	2.50[-5.07 10.07]	& 14.60  \\
                 & Mean RR	& \textbf{-8.06[-15.25 -0.88]*} &	12.76 \\
                 & RMSSD	& -1.52[-3.12 0.07]	& 2.09 \\
        \hline
        \hline
        \bottomrule
        \multicolumn{7}{p{251pt}}{Significant levels of coefficients and $F$ are highlighted in bold: $p<0.001$(***), $p<0.01$(**), and $p<0.05$(*). Adjusted $R^{2}$ and VIF are included for multiple regression models. }\\
    \end{tabular}
\label{tab2}
\end{table}

We identified the optimal feature combination in subsequent multiple regression models (models 8-12). Adding more features, i.e., incorporating acrophase from multiple sensor data, improved the model's ability to explain MEQ score variances. When using acrophase derived from the smartwatch, which included activity and heart data, model 9 ($R^2=0.54$, $F(3,32)=12.70$, $p<0.001$) had slightly superior explanatory power compared to Actigraph activities ($R^2=0.51$). When combining data from the smartwatch and the CBT sensor (model 12), acrophase from activity, temperature, and heart data were significantly associated with MEQ scores ($R^2=0.64$, $F(6,29)=8.75$, $p<0.001$). However, high multicollinearity appeared in the case of acrophase of HR-related data (VIF over 5). In the sensitivity analysis, after adjusting for age and sex in the regression models, the results remained consistent for most models except for models 4 and 7, where $R^2$ increased from 0.14 to 0.28 for SkinT and 0.13 to 0.24 for RMSSD.

\begin{figure*}[!t]
    \centering
    \includegraphics[width=0.85\textwidth]{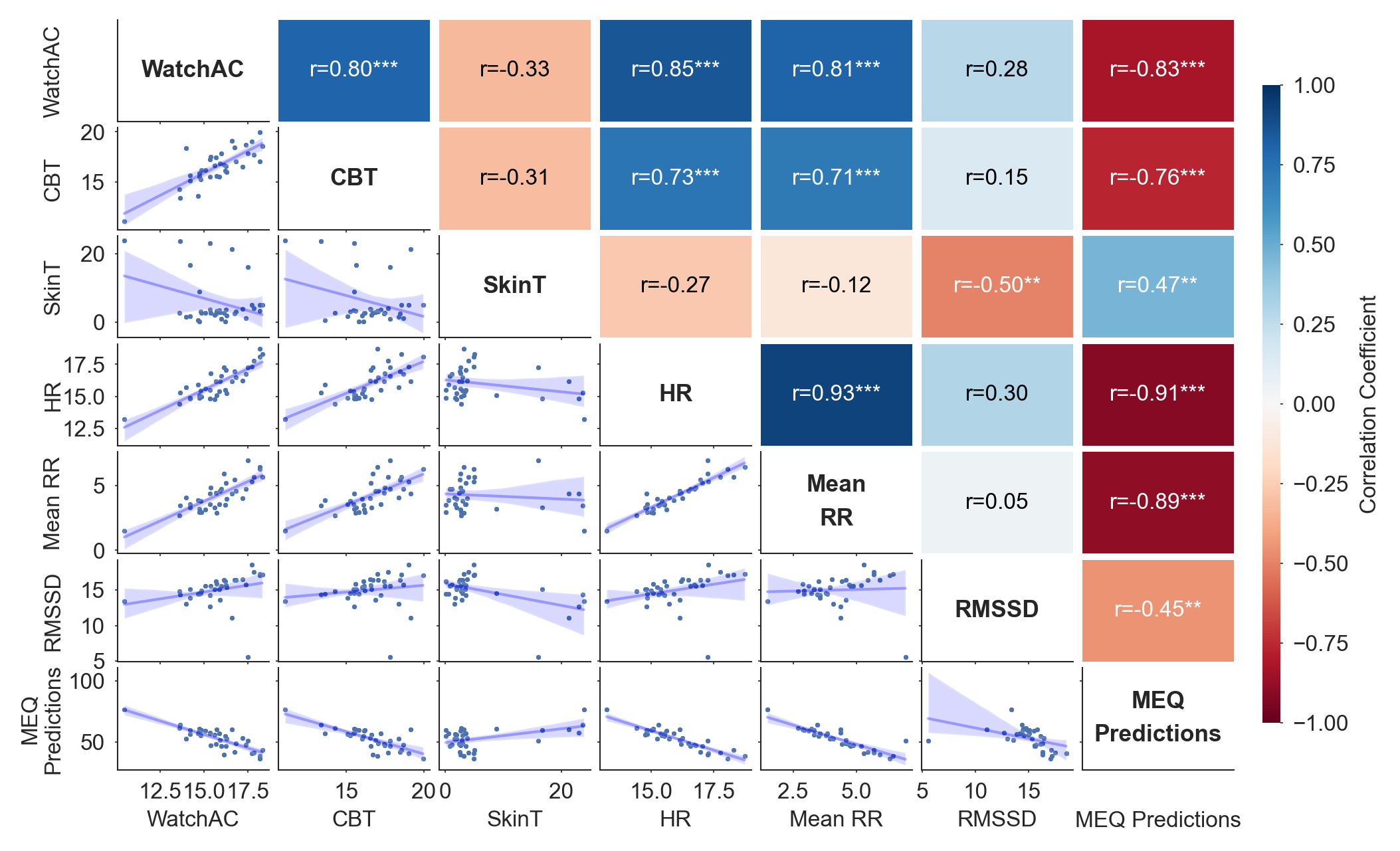}
    \caption{
    Scatter plots of acrophase and MEQ predictions, with linear regression fits (lines) and confidence intervals (shaded areas) in the bottom-left corner. Correlation coefficients in the upper-right corner. Significant levels: $p<0.001$(***) and $p<0.01$(**).}
    \label{fig3}
\end{figure*}

Taking model 12 as an illustration, Fig.~\ref{fig3} shows pairwise correlations between acrophase derived from different sensor data and MEQ predictions (N=36). We found that lower predicted MEQ scores (indicating greater eveningness) were significantly correlated with a delayed acrophase derived from HR, Mean RR, WatchAC, and CBT ($r=-0.91$, $r=-0.89$, $r=-0.83$, $r=-0.76$; all $p<0.001$). Acrophase of SkinT was the only feature exhibiting a positive relationship with MEQ predictions ($r=0.47$, $p<0.01$). In pairwise correlations among the six features, we observed significantly strong positive correlations between the acrophase of WatchAC and that of HR, Mean RR, and CBT ($r=0.85$, $r=0.81$, $r=0.80$; all $p<0.001$). Acrophase of CBT was not significantly correlated with skinT ($r=-0.31$), but it was positively correlated with WatchAC and HR ($r=0.80$, $r=0.73$; both $p<0.001$). Among all features, the acrophase of RMSSD displayed the weakest correlation with the rest features.

\subsection{Assess Acrophase Variation Across Chronotype Groups}

\begin{figure*}[!t]
    \centering
    \includegraphics[width=\textwidth]{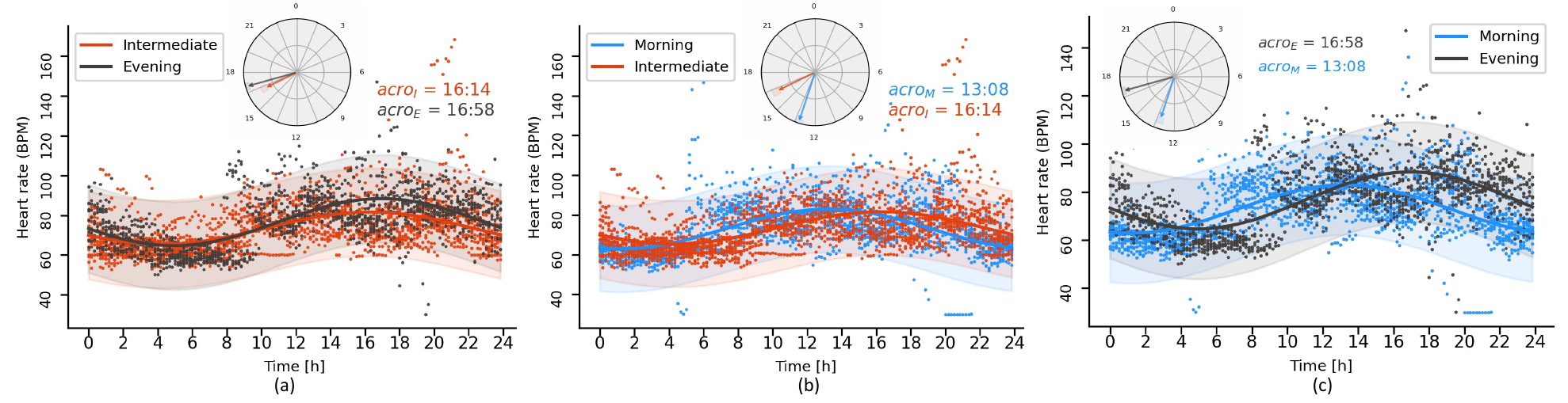}
    \caption{Acrophase comparison between Intermediate and Evening (a), Morning and Intermediate (b), and Morning and Evening (c) individuals.}
    \label{fig4}
\end{figure*}

We categorized the population into three groups based on MEQ scores: Evening (N=6, 67\% female, MEQ $36.83\pm3.83$, age $28.67\pm6.15$), Intermediate (N=16, 44\% female, MEQ $49.65\pm4.79$, age $31.94\pm6.76$), and Morning (N=14, 43\% female, MEQ $63.40\pm4.45$, age $36.14\pm13.51$). Table.~\ref{tab3} shows differences between studied acrophase among three chronotype groups. Except for skinT’s acrophase, acrophase from other sensors showed significant ($p<0.05$) differences among at least two groups. No significant ($p<0.05$) difference was observed between the Evening and Intermediate groups across all sensor data for pairwise group comparison. Nevertheless, significant differences emerged between the Intermediate and Morning groups, particularly in acrophases from HR-related data, Mean RR ($p<0.001$), HR ($p<0.01$), and activity data, ActiAC and WatchAC (both $p<0.01$). Significant differences between the Evening and Morning groups were observed in ActiAC, CBT, HR, Mean RR, RMSSD (all $p<0.01$), and WatchAC ($p<0.05$). Taking acrophase from HR as an example, Fig.~\ref{fig4} illustrates HR and its acrophase across two weeks of three individuals selected from each group. The Morning individual exhibited an acrophase at 13:08, the Intermediate at 16:14, and the Evening at 16:58.

\begin{table*}[!t]
    \centering
    \caption{Chronotype comparison based on acrophase derived from sensor data (N=36)}
    \begin{tabular}{l|ccc|ccccc}
        \hline
        \hline
        \toprule
        \textbf{Acrophase[h]} & \multicolumn{3}{|c|}{\textbf{Group descriptive statistics}} & \multicolumn{4}{c}{\textbf{Group comparison}} \\ 
         & Evening	& Intermeidate	& Morning	& KWtest	& Wtest(E-I)	& Wtest(I-M)	& Wtest(E-M)\\
        \hline
        \midrule
        \textbf{ActiAC} &	16.41(0.98)	& 16.05(1.75)	& 14.99(1.18)	& \textbf{11.75**}	& 1.00	& \textbf{2.64**} &	\textbf{3.05**}\\
        \textbf{WatchAC} &	16.43(2.02)	& 16.26(2.21)	& 14.92(1.78)	& \textbf{9.41**}	& 0.74	& \textbf{2.64**}	& \textbf{2.39*} \\
        \textbf{CBT}	& 17.95(1.26)	& 16.79(1.81)	& 15.69(1.32)	& \textbf{8.74*} 	& 1.77	& 1.77	& \textbf{2.72**}\\
        \textbf{SkinT}	& 3.12(2.09)	& 3.15(1.78)	& 2.79(5.71)	& 0.14	&-0.04	& 0.33	& 0.29\\
        \textbf{HR}	& 17.11(1.48)	& 16.53(1.73)	& 15.18(0.74)	& \textbf{12.85**}	& 1.22	& \textbf{2.79**}	& \textbf{3.09**} \\
        \textbf{Mean RR} &	5.55(1.18)	& 4.64(1.3)	& 3.39(0.58)	& \textbf{15.31***}	& 1.14	& \textbf{3.33***}	& \textbf{3.05**} \\
        \textbf{RMSSD}	& 16.34(0.68)	& 15.57(1.55)	& 14.49(1.41)	& \textbf{8.59*} &	1.03	& \textbf{2.14*} & 	\textbf{2.68**} \\
        \hline
        \hline
        \bottomrule
    \end{tabular}
    \label{tab3}
    \vspace{0.2cm}
    \parbox{\linewidth}{\raggedright Acrophase in each group is reported as medians (IQRs). E, I, and M: Evening, Intermediate, and Morning groups. KWtest: Kruskal-Wallis test; Wtest: Wilcoxon rank-sum test. Significant levels of tests are denoted in bold: $p<0.001$(***), $p<0.01$(**), and $p<0.05$(*).}
\end{table*}

\subsection{Data Transmission Rate and Device Comfort}
For all participants (N=36), the Actigraph’s data miss rate was $5.80\pm4.00\%$, range 0.72-16.29\%. In comparison, the CBT sensor displayed a slightly higher miss rate of $6.04\pm6.94\%$, 0.67-37.61\%, whereas the smartwatch demonstrated a notably higher data miss rate of $13.74\pm10.46\%$, 3.80-40.08\%. For Actigraph and smartwatches, data misses happened during water-related activities, sports with potential device damage (e.g., mountain climbing), social gatherings, and occasional arm injuries. Additionally, the smartwatch required daily charging of at least one hour. Participants occasionally overcharged it and forgot to wear it afterward. Disruptions occurred when the operating system terminated our data collection app. Initial technical problems like platform instability were addressed with a more stable data collection application during the study. CBT sensors, with 40\% being non-waterproof, experienced data misses during showers. The sensor required 2-3 hours of charging every 4-5 days, and participants had to open the app daily or before removing the sensor for data synchronization to prevent misses.

Based on comfort ratings from all participants (N=36), smartwatches scored an average of $3.38\pm1.10$, whereas CBT sensors scored $3.17\pm0.88$, indicating slightly greater comfort with smartwatches. Dissatisfaction with CBT sensors is primarily related to the attachment method, with 23 (63.89\%) participants choosing the medical patch. Of patch users, 18 (78\%) reported skin irritations, typically resolved after removal and cleaning. Three participants reported persistent irritations; one noted discomfort during sleep due to belt constriction. Despite concerns about daily charging and sleep discomfort, smartwatches were more acceptable, with 23 (64\%) participants willing to wear them daily compared to 12 (33\%) for CBT sensors.

\section{Discussion}
Our study contributed to circadian rhythm research by formally exploring and validating commercial consumer sensors compared to wrist-worn Actigraph and circadian chronotype. We conducted an observational study with 36 successful participants, with a sample size surpassing other comparative studies using consumer-grade devices \cite{lee_2017_comparison, huang_2021_predicting, smagula_2021_initial}. We simultaneously monitored activity, CBT, and HR using consumer sensors for circadian rhythm assessment. Additionally, it marks the inaugural use of the novel CBT sensor, CALERAresearch, in a two-week longitudinal circadian rhythm study. Utilizing commercial devices, our approach is scalable and easily applicable for circadian rhythm monitoring compared to standard clinical measurement or Actigraph. Using a unified method, we assessed the smartwatch's accuracy by comparing its activities with those from Actigraph. The strong agreement observed between these two devices suggested the possibility of substituting Actigraph with smartwatches. We found acrophase to be the most consistent and informative metric, whereas amplitude and mesor varied across sensors. Among non-parametric metrics, IS and IV partially aligned with Actigraph, but others performed poorly. Acrophase, derived from consumer sensors including activities, CBT, and HR-related data, showed strong correlations with Actigraph's activities and chronotype scores. Acrophase derived from the smartwatch and the CBT sensor could distinguish between Morning and Evening or Intermediate chronotypes. Additionally, we established a data collection system, allowing near real-time user data acquisition for continuous circadian rhythm monitoring. 

We observed that the HR obtained from the smartwatch exhibited the highest relationship with chronotype ($r=-0.73$), surpassing the reference Actigraph ($r=-0.71$). When predicting MEQ scores, HR's acrophase achieved the best fitness ($R^2=0.53$), comparable to Actigraph ($R^2=0.51$). Combining the CBT sensor and smartwatch yielded an enhanced MEQ prediction ($R^2=0.64$), incorporating activity, temperature, and HR-related data. This emphasizes the potential benefits of employing multiple consumer devices for circadian rhythm assessment. The stability in the sensitivity analysis, with age and sex as controls, suggests robust associations not significantly influenced by demographic factors. 

Prior studies have predominantly employed acrophase metrics derived from Actigraph data. In our approach, we have extracted the acrophase from data gathered through commercial devices, attaining comparable results. Additionally, our findings are consistent with existing research that correlates Actigraph-derived metrics with questionnaire-based evaluations. Schneider et~al. showed a correlation between Actigraphy acrophase and MEQ scores of 0.62~\cite{schneider_2021_human}. Kaufmann et~al. revealed a significant correlation between the daytime midpoint mean and the Composite Scale of Morningness questionnaire ($r=-0.56$)~\cite{kaufmann_2018_daytime}. Roveda et~al. found a strong and inverse association between MEQ and actigraphy-based acrophase ($r=-0.84$)~\cite{roveda_2017_predicting}. 

In assessing the practical use of these commercial sensors, user commitment was crucial for successful data collection. Real-time data access and a user-friendly interface enhanced participant experience and willingness to wear the devices. CALERAresearch surpassed the smartwatch in data collection rates, but problems in wearing comfort and active engagement necessitate a broader acceptance solution. Despite challenges in charging and data collection stability, the Galaxy Watch 5 emerged as a promising alternative for circadian rhythm studies due to higher willingness to wear and superior chronotype prediction. Galaxy Watch's data collection rate outperformed other comparable studies like the Apple Watch, which experienced data loss around 6-8 hours every 1-2 days~\cite{huang_2021_predicting}. The reference Actigraph had no data misses due to charging within two weeks but might face challenges in more extended studies due to its 25-day battery life at a 30Hz sampling rate. Despite more data loss in tested commercial devices, circadian rhythm metrics from these devices remained comparable to Actigraph. 

Although our study assessed the commercial wearables in both technical and practical prospective, we acknowledge its limitations. Melatonin was not referenced due to the lack of a validated at-home measurement method. The participant composition, with many recruited from universities, may only partially represent larger populations. Enhancing diversity would improve the generalizability of our findings. We observed data loss for various reasons, including active charging, app engagement, and platform stability. Charging problems can potentially be addressed with the new Galaxy Watch Pro, which has an extended battery life. Concurrently, we persist in improving the stability of our CLAID platform and standardizing data upload processes, eliminating the need for manual data sync. Last, as we mainly focused on collecting data, we did not guide participants on adjusting rhythms based on their data input. Recent studies indicate that disruptions in circadian rhythm, as estimated through wearables, correlate with reduced health span and accelerated biological aging, highlighting the pivotal role of circadian rhythm in promoting healthy longevity~\cite{shim_2023_wearablebased}. This underscores the substantial potential of digital devices in optimizing individual rhythmicity and enhancing circadian health. Future intervention research is warranted to utilize consumer devices to assess a personalized circadian rhythm and optimize circadian health for an extended period.

Author's photographs and biographies not available at the time of publication.

\bibliography{Bibliography}

\section*{References}
\begin{thebibliography}{00}\leftskip1pc

\bibitem{vitaterna_2001_overview}
M.~H. Vitaterna, J.~S. Takahashi, and F.~W. Turek, ``Overview of circadian rhythms,'' \emph{Alcohol Research \& Health}, vol.~25, p. 85–93, 2001. [Online]. Available: \url{https://www.ncbi.nlm.nih.gov/pmc/articles/PMC6707128/}

\bibitem{kim_2013_circadian}
M.~J. Kim, J.~H. Lee, and J.~F. Duffy, ``Circadian rhythm sleep disorders,'' \emph{Journal of clinical outcomes management: JCOM}, vol.~20, p. 513–528, 11 2013. [Online]. Available: \url{https://pubmed.ncbi.nlm.nih.gov/25368503/}

\bibitem{cappuccio_2010_sleep}
 
F.~P. Cappuccio, L.~D'Elia, P.~Strazzullo, and M.~A. Miller, ``Sleep duration and all-cause mortality: A systematic review and meta-analysis of prospective studies,'' \emph{Sleep}, vol.~33, pp. 585--592, 05 2010. [Online]. Available: \url{https://www.ncbi.nlm.nih.gov/pmc/articles/PMC2864873/}
  

\bibitem{cappuccio_2011_sleep}
F.~P. Cappuccio, D.~Cooper, L.~D'Elia, P.~Strazzullo, and M.~A. Miller, ``Sleep duration predicts cardiovascular outcomes: a systematic review and meta-analysis of prospective studies,'' \emph{European Heart Journal}, vol.~32, pp. 1484--1492, 02 2011.

\bibitem{cappuccio_2009_quantity}
F.~P. Cappuccio, L.~D'Elia, P.~Strazzullo, and M.~A. Miller, ``Quantity and quality of sleep and incidence of type 2 diabetes: A systematic review and meta-analysis,'' \emph{Diabetes Care}, vol.~33, pp. 414--420, 11 2009.

\bibitem{cappuccio_2008_metaanalysis}
 
F.~P. Cappuccio, F.~M. Taggart, N.-B. Kandala, A.~Currie, E.~Peile, S.~Stranges, and M.~A. Miller, ``Meta-analysis of short sleep duration and obesity in children and adults,'' \emph{Sleep}, vol.~31, pp. 619--26, 2008. [Online]. Available: \url{https://www.ncbi.nlm.nih.gov/pmc/articles/PMC2398753/}
  

\bibitem{rahman_2022_dynamic}
S.~A. Rahman, M.~A. St.~Hilaire, L.~K. Grant, L.~K. Barger, G.~C. Brainard, C.~A. Czeisler, E.~B. Klerman, and S.~W. Lockley, ``Dynamic lighting schedules to facilitate circadian adaptation to shifted timing of sleep and wake,'' \emph{Journal of Pineal Research}, vol.~73, 05 2022.

\bibitem{reid_2019_assessment}
 
K.~J. Reid, ``Assessment of circadian rhythms,'' \emph{Neurologic clinics}, vol.~37, p. 505–526, 08 2019. [Online]. Available: \url{https://www.ncbi.nlm.nih.gov/pmc/articles/PMC6857846/}
  

\bibitem{ancoliisrael_2003_the}
 
S.~Ancoli-Israel, R.~Cole, C.~Alessi, M.~Chambers, W.~Moorcroft, and C.~P. Pollak, ``The role of actigraphy in the study of sleep and circadian rhythms,'' \emph{Sleep}, vol.~26, pp. 342--392, 05 2003. [Online]. Available: \url{https://aasm.org/resources/practicereviews/cpr_actigraphy.pdf}
  

\bibitem{ali_2023_circadian}
 
F.~Z. Ali, R.~V. Parsey, S.~Lin, J.~Schwartz, and C.~DeLorenzo, ``Circadian rhythm biomarker from wearable device data is related to concurrent antidepressant treatment response,'' \emph{npj Digital Medicine}, vol.~6, p. 1–11, 04 2023. [Online]. Available: \url{https://www.nature.com/articles/s41746-023-00827-6}
  

\bibitem{horne_1976_a}
 
J.~A. Horne and O.~Ostberg, ``A self-assessment questionnaire to determine morningness-eveningness in human circadian rhythms,'' \emph{International journal of chronobiology}, vol.~4, p. 97–110, 01 1976. [Online]. Available: \url{https://europepmc.org/article/med/1027738}
  

\bibitem{taillard_2004_validation}
J.~Taillard, P.~Philip, J.-F. Chastang, and B.~Bioulac, ``Validation of horne and ostberg morningness-eveningness questionnaire in a middle-aged population of french workers,'' \emph{Journal of Biological Rhythms}, vol.~19, pp. 76--86, 02 2004.

\bibitem{suarez_2021_circadian}
A.~Suarez, F.~Nunez, and M.~Rodriguez-Fernandez, ``Circadian phase prediction from non-intrusive and ambulatory physiological data,'' \emph{IEEE Journal of Biomedical and Health Informatics}, vol.~25, pp. 1561--1571, 05 2021.

\bibitem{lee_2017_comparison}
 
H.-A. Lee, H.-J. Lee, J.-H. Moon, T.~Lee, M.-G. Kim, H.~In, C.-H. Cho, and L.~Kim, ``Comparison of wearable activity tracker with actigraphy for sleep evaluation and circadian rest-activity rhythm measurement in healthy young adults,'' \emph{Psychiatry Investigation}, vol.~14, p. 179, 2017. [Online]. Available: \url{https://www.ncbi.nlm.nih.gov/pmc/articles/PMC5355016/}
  

\bibitem{huang_2021_predicting}
Y.~Huang, C.~Mayer, P.~Cheng, A.~Siddula, H.~J. Burgess, C.~Drake, C.~Goldstein, O.~Walch, and D.~B. Forger, ``Predicting circadian phase across populations: a comparison of mathematical models and wearable devices,'' \emph{Sleep}, vol.~44, 05 2021.

\bibitem{smagula_2021_initial}
S.~F. Smagula, S.~T. Stahl, R.~T. Krafty, and D.~J. Buysse, ``Initial proof of concept that a consumer wearable can be used for real-time rest-activity rhythm monitoring,'' \emph{SLEEP}, vol.~45, 12 2021.

\bibitem{vanmarkenlichtenbelt_2006_evaluation}
 
W.~D. van Marken~Lichtenbelt, H.~A.~M. Daanen, L.~Wouters, R.~Fronczek, R.~J. E.~M. Raymann, N.~M.~W. Severens, and E.~J.~W. Van~Someren, ``Evaluation of wireless determination of skin temperature using ibuttons,'' \emph{Physiology \& Behavior}, vol.~88, p. 489–497, 07 2006. [Online]. Available: \url{https://www.sciencedirect.com/science/article/pii/S0031938406001818}
  

\bibitem{huang_2021_distinct}
Y.~Huang, C.~Mayer, O.~J. Walch, C.~Bowman, S.~Sen, C.~Goldstein, J.~Tyler, and D.~B. Forger, ``Distinct circadian assessments from wearable data reveal social distancing promoted internal desynchrony between circadian markers,'' \emph{Frontiers in Digital Health}, vol.~3, 11 2021.



\bibitem{ajevi_2022_a}
M.~Ajčević, A.~Buoite~Stella, G.~Furlanis, P.~Caruso, M.~Naccarato, A.~Accardo, and P.~Manganotti, ``A novel non-invasive thermometer for continuous core body temperature: Comparison with tympanic temperature in an acute stroke clinical setting,'' \emph{Sensors}, vol.~22, p. 4760, 06 2022.

\bibitem{rerabek_2022_circadian}
 
M.~Rerabek, G.~Schiboni, L.~Durrer, R.~Oliveras, P.~Eib, F.~Rouchat, A.~Probst, M.~Schmidt, and K.~Kryszczuk, ``Circadian rhythm tracking using core body temperature estimates from wearable sensor data,'' digitalcollection.zhaw.ch, 2022. [Online]. Available: \url{https://digitalcollection.zhaw.ch/handle/11475/26865}
  

\bibitem{flora_2021_highfrequency}
C.~Flora, J.~Tyler, C.~Mayer, D.~E. Warner, S.~N. Khan, V.~Gupta, R.~Lindstrom, A.~Mazzoli, M.~Rozwadowski, T.~M. Braun, M.~Ghosh, D.~B. Forger, S.~W. Choi, and M.~Tewari, ``High-frequency temperature monitoring for early detection of febrile adverse events in patients with cancer,'' \emph{Cancer Cell}, vol.~39, pp. 1167--1168, 09 2021.

\bibitem{langer_2023_claid}
P.~Langer, E.~Fleisch, and F.~Barata, ``Claid: Closing the loop on ai \& data collection -- a cross-platform transparent computing middleware framework for smart edge-cloud and digital biomarker applications,'' \emph{arXiv (Cornell University)}, 10 2023.

\bibitem{neishabouri_2022_quantification}
 
A.~Neishabouri, J.~Nguyen, J.~Samuelsson, T.~Guthrie, M.~Biggs, J.~Wyatt, D.~Cross, M.~Karas, J.~H. Migueles, S.~Khan, and C.~C. Guo, ``Quantification of acceleration as activity counts in actigraph wearable,'' \emph{Scientific Reports}, vol.~12, p. 11958, 07 2022. [Online]. Available: \url{https://www.nature.com/articles/s41598-022-16003-x}
  

\bibitem{bate_2023_the}
G.~L. Bate, C.~Kirk, R.~Zia, Y.~Guan, A.~J. Yarnall, S.~D. Din, and R.~A. Lawson, ``The role of wearable sensors to monitor physical activity and sleep patterns in older adult inpatients: A structured review,'' \emph{Sensors}, vol.~23, pp. 4881--4881, 05 2023.

\bibitem{dawes_1972_circadian}
C.~Dawes, ``Circadian rhythms in human salivary flow rate and composition,'' \emph{The Journal of Physiology}, vol. 220, pp. 529--545, 02 1972.

\bibitem{spengler_2000_endogenous}
C.~SPENGLER and S.~SHEA, ``Endogenous circadian rhythm of pulmonary function in healthy humans,'' \emph{American Journal of Respiratory and Critical Care Medicine}, vol. 162, pp. 1038--1046, 09 2000.

\bibitem{mokon_2020_cosinorpy}
M.~Moškon, ``Cosinorpy: a python package for cosinor-based rhythmometry,'' \emph{BMC Bioinformatics}, vol.~21, 10 2020.

\bibitem{fossion_2018_quantification}
 
R.~Fossion, A.~L. Rivera, J.~C. Toledo-Roy, and M.~Angelova, ``Quantification of irregular rhythms in chronobiology: A time- series perspective,'' www.intechopen.com, 07 2018. [Online]. Available: \url{https://www.intechopen.com/chapters/60261}
  

\bibitem{ticher_1994_human}
A.~Ticher, L.~Sackett-Lundeen, I.~E. Ashkenazi, and E.~Haus, ``Human circadian time structure in subjects of different gender and age,'' \emph{Chronobiology International}, vol.~11, pp. 349--355, 01 1994.

\bibitem{zacharia_2014_development}
T.~Zacharia, J.~James, H.~Prakash, R.~T. Mohan, and B.~Rajashekhar, ``Development and standardization of morningness-eveningness questionnaire (meq) in the indian language kannada,'' \emph{The International Tinnitus Journal}, vol.~19, 2014.

\bibitem{roveda_2017_predicting}
E.~Roveda, J.~Vitale, A.~Montaruli, L.~Galasso, F.~Carandente, and A.~Caumo, ``Predicting the actigraphy-based acrophase using the morningness–eveningness questionnaire (meq) in college students of north italy,'' \emph{Chronobiology International}, vol.~34, pp. 551--562, 02 2017.

\bibitem{schneider_2021_human}
J.~Schneider, E.~Fárková, and E.~Bakštein, ``Human chronotype: Comparison of questionnaires and wrist-worn actigraphy,'' \emph{Chronobiology International}, vol.~39, pp. 205--220, 11 2021.

\bibitem{kaufmann_2018_daytime}
C.~N. Kaufmann, A.~Gershon, C.~A. Depp, S.~Miller, J.~M. Zeitzer, and T.~A. Ketter, ``Daytime midpoint as a digital biomarker for chronotype in bipolar disorder,'' \emph{Journal of Affective Disorders}, vol. 241, pp. 586--591, 12 2018.

\bibitem{shim_2023_wearablebased}
 
J.~Shim, E.~Fleisch, and F.~Barata, ``Wearable-based accelerometer activity profile as digital biomarker of inflammation, biological age, and mortality using hierarchical clustering analysis in nhanes 2011–2014,'' \emph{Scientific Reports}, vol.~13, p. 9326, 06 2023. [Online]. Available: \url{https://www.nature.com/articles/s41598-023-36062-y}
  


\end{thebibliography}

\end{document}